\documentclass[a4paper,one]{amsart}
\usepackage{amsthm}
\usepackage{amsmath}
\usepackage{amsfonts}
\usepackage{amssymb}
\usepackage{mathrsfs}
\usepackage{hyperref}
\usepackage[numbers]{natbib}
\usepackage{fancyhdr}
\theoremstyle{plain}
\newtheorem{lem}{Lemma}

\newtheorem{thm}[lem]{Theorem}

\theoremstyle{definition}

\newtheorem{rem}[lem]{Remark}
\newcommand{\R}{\mathbb R}
\newcommand{\Z}{\mathbb Z}
\newcommand{\N}{\mathbb N}
\newcommand{\Diff}{\mbox{\rm Diff}}

\newcommand{\id}{\mbox{\rm id}}
\newcommand{\dx}{\,\text{\rm d}x}
\renewcommand{\d}{\,\text{\rm d}}

\renewcommand{\S}{\mathbb S}
\renewcommand{\phi}{\varphi}
\newcommand{\muDP}{\mu\text{\rm DP}}
\newcommand{\muCH}{\mu\text{\rm CH}}

\newcommand{\norm}[1]{\left|\!\left|#1\right|\!\right|}

\newcommand{\ska}[2]{\left\langle #1,#2\right\rangle}
\newcommand{\set}[2]{\left\{#1;\;#2\right\}}
\newcommand{\bea}{\begin{eqnarray}}
\newcommand{\eea}{\end{eqnarray}}
\newcommand{\beq}{\begin{equation}}
\newcommand{\eeq}{\end{equation}}
\begin{document}
\title{Global existence and blow-up for a weakly dissipative $\mu$DP equation}
\author{Martin Kohlmann}
\address{Institute for Applied Mathematics, University of Hannover, D-30167 Hannover, Germany}
\email{kohlmann@ifam.uni-hannover.de}
\date{\today}
\begin{abstract} In this paper, we study a weakly dissipative variant
of the periodic Degasperis-Procesi equation. We show the local well-posedness of the associated Cauchy
problem in $H^s(\S)$, $s>3/2$, and discuss the precise blow-up
scenario for $s=3$. We also present explicit examples for globally existing
solutions and blow-up.
\end{abstract}
%
%
%
%
\maketitle
\section{Introduction}
In recent years, the family
\bea\label{beqn}y_t=-(y_xu+bu_xy),\quad y=u-u_{xx},\eea
of nonlinear equations has been studied extensively; see
\cite{EK,EY}. Here, $u(t,x)$ depends on a time variable $t\geq 0$ and
a space variable $x$ with $x\in\S\simeq\R/\Z$ for the periodic
equation and $x\in\R$ in the non-periodic case. Equation
(\ref{beqn}) is called $b$-equation and (\ref{beqn}) becomes the
Camassa-Holm (CH) equation for $b=2$ and the Degasperis-Procesi
(DP) equation if $b=3$. Moreover, the corresponding family of
$\mu$-equations, where
$$y=\mu(u)-u_{xx},\quad\mu(u)=\int u(t,x)\dx$$
in (\ref{beqn}), has been studied, e.g., in \cite{FLQ,LMT}. It is
known that the $b$-equation models the unidirectional motion of 1D
water waves over a flat bed; for the hydrodynamical relevance we
refer to, e.g., \cite{CH,DHH} and \cite{CL,DGH,J}. It turned out
that the $b$-equation is integrable only if $b\in\{2,3\}$ and in
\cite{LMT} the authors mention that a similar result is
conjectured for the family of $\mu$-equations. The Cauchy problems
for the $b$-equation and its $\mu$-variant have been discussed in
\cite{EY,LMT}. In particular, for $b\in\{2,3\}$, local
well-posedness results in the periodic and in the real line-case
as well as blow-up and criteria for the global existence of strong and
weak solutions have been established; see, e.g., \cite{CE,ELY07,Yi}. In
addition, CH and DP and $\muCH$ and $\muDP$ admit peaked
solitons, which make them attractive among the integrable
equations; cf.~\cite{CH,DHH,LMT}.\\
\indent In general, it is difficult to avoid energy dissipation
mechanisms in the modeling of fluids. Ott and Sudan \cite{OS}
discussed the KdV equation under the influence of energy
dissipation. Ghidaglia \cite{G} studied the behavior of solutions
of the weakly dissipative KdV equation as a finite-dimensional
dynamical system. Some results for the weakly dissipative CH
equation are proved in \cite{WYi} and recently, \cite{EWY,Y}
discussed blow-up and global existence for the
weakly dissipative DP equation.\\
\indent The goal of the present paper is to study the Cauchy
problem for the periodic weakly dissipative $\muDP$ equation
\bea\label{wdmuDP}\left\{\begin{array}{rcl}
  y_t+uy_x+3u_xy+\lambda y & = & 0, \\
  y & = & \mu(u)-u_{xx}, \\
  u(0,x) & = & u_0(x). \\
\end{array}\right.\eea
Here the function $u(t,x)$ is depending on time $t\geq 0$ and a
space variable $x\in\S$ and $\mu$ is the projection
$\mu(u)=\int_0^1u(t,x)\dx$. The constant $\lambda$ is assumed to
be positive and the term $\lambda(\mu(u)-u_{xx})$ models energy
dissipation. By the replacement $\mu(u)\mapsto u$ in
(\ref{wdmuDP}), we obtain the weakly dissipative DP equation. Note
that the quantity $E_1(u)=\int_\S y\d x$ is conserved for the DP
equation and that $E_1$ can be interpreted as an energy, since it
equals (up to a factor) a Hamiltonian function for the DP as
explained in \cite{LMT}. However, for the weakly dissipative DP
equation, $\frac{\text{d}}{\text{d}t}E_1(u)=-\lambda\mu(u)$, such
that $\mu(u_0)>0$ implies that the wave's energy decreases
as $t$ increases. The weak dissipation also breaks other
conservation laws of the DP equation like $E_2(u)=\int_\S yv\dx$
or $E_3(u)=\int_\S u^3\dx$, where $v=(4-\partial_x^2)^{-1}u$; cf.
\cite{WY}.\\
\indent The general framework in which we discuss equation
(\ref{wdmuDP}) is based on geometric and Lie theoretic techniques
as introduced in \cite{A,EM}. Equation (\ref{wdmuDP}) can be
regarded as an evolution equation on the group $\Diff^s(\S)$ of
orientation-preserving $H^s$ diffeomorphisms of the circle
$\S$, for $s>3/2$:\ The vector field
$u(t,\cdot)\in H^s(\S)$ has a unique local flow
$\phi(t,\cdot)\in\Diff^s(\S)$ such that $\phi_t\circ\phi^{-1}=u$,
$\phi(0)=\id$ and $\phi_{tt}=-F(\phi,\phi_t)$ with some map $F$
defined on $\Diff^s(\S)\times H^s(\S)$. This equation can be
handled with standard ODE methods for Banach spaces. Altogether,
it will turn out that the weakly dissipative $\muDP$ equation
behaves quite similarly to the $\muDP$ equation (for which
$\lambda=0$) or the weakly dissipative
DP equation.\\
\indent The paper is organized as follows: In Section
\ref{sec_LWP}, we prove local well-posedness for the initial value
problem (\ref{wdmuDP}) with $u_0\in H^s(\S)$ for $s>3/2$. In
Section \ref{sec_GWP_blowup}, we show that for smooth initial data
with zero mean, the solution $u(t,\cdot)$ of (\ref{wdmuDP}) can
blow up in finite time. If $\mu(u_0)\neq 0$ and $\mu(u_0)-u_{0xx}$
is non-negative or non-positive, the corresponding solution
$u(t,\cdot)$ will exist globally in time.\\[.25cm]
\noindent\emph{Acknowledgement.} The author thanks the anonymous referee for useful remarks which helped to improve the structure of the paper.
\section{Local well-posedness}\label{sec_LWP}
In this section, we aim to establish a local well-posedness result
for the Cauchy problem (\ref{wdmuDP}). The proof uses some
geometric arguments and is based on a reformulation of the weakly
dissipative $\muDP$ as a quasi-linear evolution equation; cf.
\cite{LMT}.
We will use the notation
$$\Lambda_{\mu}^2:=\mu-\partial_x^2$$
and write $y=\Lambda_\mu^2u$, $y_0=\Lambda_\mu^2u_0$. It is not
hard to see that $\Lambda_{\mu}^2$ is a topological isomorphism
between the Sobolev spaces $H^s(\S)$ and $H^{s-2}(\S)$, $s\geq 2$,
cf.~Sect.~\ref{appendix}; the inverse of $\Lambda_\mu^2$ is
denoted by $\Lambda_{\mu}^{-2}$. In the following, we are only
interested in Sobolev functions of class $s=3$, but our first
theorem deals with the more general case $s>3/2$. Note that
$H^s(\S)\subset C^{k}(\S)$, $s>k+1/2$, and that the square of
$\norm{\cdot}_s=\norm{\cdot}_{H^s(\S)}$ is the quadratic form
(with respect to the $L_2$ inner product) induced by the operator
$Q^{2s}=(1-\partial_x^2)^s$, which also defines an isomorphism
between the spaces $H^k(\S)$ and $H^{k-2s}(\S)$. The group of
orientation-preserving circle diffeomorphisms $\S\to\S$ of class
$H^s$ is denoted by $\Diff^s(\S)$, i.e.,
$$\Diff^s(\S):=\set{\phi\in H^s(\S)}{\phi\text{ is bijective, orientation-preserving and }\phi^{-1}\in H^s(\S)}.$$
Observe that $T_{\phi}\Diff^s(\S)\simeq H^s(\S)$ for any
$\phi\in\Diff^s(\S)$. The following lemma establishes that
$\Diff^s(\S)$ is a topological group for $s>3/2$. The reader can
find a proof in \cite{M}.
\lem\label{composition} For $s>3/2$, the composition map
$\phi\mapsto\omega\circ\phi$ with an $H^s$ function $\omega$ and the
inversion map $\phi\mapsto\phi^{-1}$ are continuous maps
$\Diff^s(\S)\to H^s(\S)$ and $\Diff^s(\S)\to\Diff^s(\S)$
respectively and
$$\norm{\omega\circ\phi}_{H^s}\leq C(1+\norm{\phi}_{H^s}^s)\norm{\omega}_{H^s};$$
$C$ only depending on $\sup_{x\in\S}|\phi_x(x)|$ and
$\inf_{x\in\S}|\phi_x(x)|$.
\endlem\rm
Before we proceed, we state the following lemma which has an
interesting geometric interpretation and is derived directly from the local
existence and uniqueness theorem for differential equations in Banach spaces; cf.~\cite{L}.
\lem\label{lem_flow} Let $u(t,x)$ be a time-dependent $H^s$ function on the circle for $s>3/2$. Then the problem
\bea\left\{\begin{array}{ccl}
\phi_t(t,x)&=&u(t,\phi(t,x)),\nonumber\\
\phi(0,x)&=&x,\end{array}\right.\nonumber
\eea
for $x\in\S$ and $t\geq 0$, has a unique solution $\phi\in
C^{1}([0,T_{\max}),\Diff^s(\S))$, where $T_{\max}>0$ is maximal.
\endlem\rm
\rem Note that the geometric interpretation of this lemma is that
$u(t,\cdot)$ can be regarded as a vector field on the sphere $\S$
for which we have a local flow $\phi\in\Diff^s(\S)$. Local flows
have approved to be powerful tools in the analysis of model
equations for 1D water waves, see \cite{A,EM,EK,Len}.
\endrem
In many texts, local well-posedness results for Cauchy problems
similar to (\ref{wdmuDP}) are obtained by applying Kato's theory
for abstract quasi-linear evolution equations. We now present a
method of proof which is based on a geometric argument, most
importantly using local flows as introduced in the above lemma. A
technical disadvantage of this method is that it does not yield a
priori a maximal existence time for our solution which we will
obtain inductively.
The key idea is to rewrite the weakly dissipative $\muDP$ equation
in the form
\bea u_t+uu_x+3\mu(u)\partial_x\Lambda_{\mu}^{-2}u+\lambda
u=0;\label{eqn_rewritten}\eea
this equation is suitable for a reformulation of
\eqref{wdmuDP} in the geometric picture, i.e.,
in terms of a local flow on the group $\Diff^s(\S)$. For the following well-posedness proof, our next lemma will play a key role.
The explicit calculations already occur in the proof of Theorem 5.1 in \cite{LMT}.
\lem\label{lem_computations} Let $R_\phi$ denote the right translation map on
$\Diff^s(\S)$ and let
$\Lambda_{\mu,\phi}^{-2}=R_\phi\circ\Lambda_{\mu}^{-2}\circ
R_{\phi^{-1}}$ and
$\partial_{x,\phi}=R_\phi\circ\partial_{x}\circ
R_{\phi^{-1}}$. Then,
\bea\label{5.9}
3\mu(\xi\circ\phi^{-1})\left(\Lambda_\mu^{-2}\partial_x
\left(\xi\circ\phi^{-1}\right)\right)\circ\phi=\Lambda_{\mu,\phi}^{-2}\partial_{x,\phi}h(\phi,\xi)\eea
for $h(\phi,\xi)=3\xi\int_\S\xi\circ\phi^{-1}\dx$.
Furthermore, we have the identities
\bea\label{5.10}\partial_\phi\Lambda_{\mu,\phi}^{-2}(v)&=&-\Lambda_{\mu,\phi}^{-2}
\left[(v\circ\phi^{-1})\partial_x,\Lambda_\mu^2\right]_\phi\Lambda_{\mu,\phi}^{-2},\\
\label{5.11}\partial_\phi\partial_{x,\phi}(v)&=&\left[(v\circ\phi^{-1})\partial_x,\partial_x\right]_\phi,\\
\label{5.12}\partial_\phi
h(\phi,\xi)(v)&=&3\xi\int_\S\xi\circ\phi^{-1}\partial_x(v\circ\phi^{-1})\dx.
\eea
\endlem\rm
Our main theorem in this section reads as follows.
\thm\label{LWP} Let $s>3/2$ and $u_0\in H^s(\S)$. Then there is
a maximal time $T\in(0,\infty]$ and a unique solution
$$u\in C\left([0,T);H^s(\S)\right)\cap C^1\left([0,T);H^{s-1}(\S)\right)$$
of the Cauchy problem \text{\rm(\ref{wdmuDP})} which depends
continuously on the initial data $u_0$, i.e., the mapping
$$H^s(\S)\to C\left([0,T);H^s(\S)\right)\cap C^1
\left([0,T);H^{s-1}(\S)\right),\quad u_0\mapsto u(\cdot,u_0)$$
is continuous.
\endthm
\proof
Writing
$$Au=3\mu(u)\partial_x\Lambda_{\mu}^{-2}u+\lambda u=A_0u+\lambda u,$$
Eq.~\eqref{eqn_rewritten} shows that (\ref{wdmuDP}) is equivalent to
$u_t+uu_x=-Au$. Let $\phi\in\Diff^s(\S)$ denote the local flow for the vector
field $u(t,\cdot)$ according to Lemma~\ref{lem_flow}, i.e., $\phi(t)$ is defined on some maximal interval $[0,T_{\max})$ and is $C^1$, with
$u\circ\phi=\phi_t$ and $\phi(0)=\id$. If we can differentiate $\phi_t$ once again, we may derive the identity
$$\phi_{tt}=(u_t+uu_x)\circ\phi=-A(\phi_t\circ\phi^{-1})\circ\phi.$$
Let $F(\phi,\phi_t):=R_{\phi}\circ A\circ R_{\phi^{-1}}\phi_t$ so
that
\bea\label{ODE}\phi_{tt}=-F(\phi,\phi_t),\quad\phi_t(0)=u_0,\quad\phi(0)=\id,\eea
which is an ordinary second order initial value problem. Interestingly, any solution $\phi$ to the initial value problem \eqref{ODE} yields a solution $u=\phi_t\circ\phi^{-1}$ to the weakly dissipative $\mu$DP equation in its initial form, with the desired regularity properties. This motivates to study the Cauchy problem for the periodic weakly dissipative $\mu$DP in the reformulation \eqref{ODE} on the diffeomorphism group of the circle.

We now decompose $F=F_1+F_2$ with $F_1=R_{\phi}\circ A_0\circ
R_{\phi^{-1}}$ and $F_2$ just being multiplication with $\lambda$. Both, $F_1$ and $F_2$ are Fr\'echet differentiable in a neighborhood of any $(\phi,\xi)\in
T\Diff^s(\S)\simeq\Diff^s(\S)\times H^s(\S)$ and the directional derivatives $\partial_{\phi}F_i$, $\partial_{\xi}F_i$, $i=1,2$, are bounded linear operators on
$H^s(\S)$ with continuous dependence on $(\phi,\xi)$; this is trivial for $F_2$ and proved in \cite{LMT} for $F_1$ by applying Lemma~\ref{lem_computations}.

Since $F$ is continuously differentiable near $(\id,0)$, the standard local
existence theorem for Banach spaces (cf.\ \cite{L}) establishes
the local well-posedness of (\ref{ODE}), i.e., there is a time
$T_1>0$ and a unique solution $(\phi,\phi_t)$ of
(\ref{ODE}) on $[0,T_1]$ with continuous dependence on $t$ and
$u_0$. To show that there is a maximal interval of existence, we
apply the Cauchy-Lipschitz Theorem once more to the problem
$\phi_{tt}=-F(\phi,\phi_t)$ with initial data
$(\phi(T_1),\phi_t(T_1))$ to continue the solution $(\phi,\phi_t)$
to a solution on a time interval $[0,T_2]$ with
$T_1<T_2$. Iterating this procedure, we obtain a monotonically
increasing sequence $(T_n)_{n\in\N}$ and an associated sequence of solutions
$$\phi_n\in C^2\left([0,T_n];\Diff^s(\S)\right),\quad\phi_{n+1}|_{[0,T_n]}=\phi_n.$$
If $(T_n)_{n\in\N}$ is bounded, $T_n\to T$ as $n\to\infty$, with a real number $T$; otherwise, $T_n\to\infty$.

Now the well-posedness of the problem (\ref{wdmuDP}) is a simple consequence of the relations
$u=\phi_t\circ\phi^{-1}$ and $u_t=-uu_x-Au$ and the fact that $\Diff^s(\S)$ is a
topological group whenever $s>3/2$.\endproof
\rem As explained in the proof of Theorem \ref{LWP}, let $(T_n)_{n\in\N}$ denote the strictly increasing sequence describing the continuation of our solution $u(t,x)$ in $H^s(\S)$. If $(T_n)_{n\in\N}$ is bounded, we say that the solution $u$ has a \emph{finite} existence time, where $T_n\to\infty$ means that the solution exists \emph{globally} in time. It is an interesting problem and the aim of the following sections to describe the behavior of finite-time solutions as
$t\to T$ from below and to find criteria for the global existence of strong solutions as well as so-called \emph{finite-time blow-up}.
\endrem
\section{Global well-posedness and blow-up}\label{sec_GWP_blowup}
In physics, a breaking wave is a wave whose amplitude reaches a
critical level at which some process can suddenly start to occur
that causes large amounts of wave energy to be transformed in
turbulent kinetic energy. At this point, simple physical models
describing the dynamics of the wave will often become invalid,
particularly those which assume linear behavior. Wave breaking has
been studied for various classes of non-linear model equations for
1D water waves and a reasonable way is to show that there is a
finite-time solution $u$ satisfying an $L_\infty$-bound for all
$t\in[0,T)$ so that the norm of $u$ is unbounded as $t\to T$ if
and only if the first order derivative $u_x$ approaches $-\infty$
as $t\to T$ from below (cf., e.g., \cite{EWY} for a discussion of
the DP equation with a dissipative term). The physical
interpretation then is that the wave steepens, while the height of
its crests stays bounded, until wave breaking occurs in the sense
that $u$ ceases to be a classical solution.
\\
\indent In this section, we first describe the blow-up of finite-time solutions of (\ref{wdmuDP}) in terms of the first order
derivative and then discuss precise blow-up settings.
Here and in what follows, we will restrict ourselves to $s=3$.
Recall that $H^3(\S)$-functions are of class $C^2$ such that there
will be no boundary terms when performing integration by parts.
\thm\label{ableitung} Given $u_0\in H^3(\S)$, the solution $u$ of
\text{\rm(\ref{wdmuDP})} obtained in Theorem \text{\rm\ref{LWP}}
blows up in finite time $T>0$ if and only if
$$\liminf_{t\to T}\min_{x\in\S} u_x(t,x)=-\infty.$$
\endthm\rm
\proof Let $T>0$ be the maximal time of existence of the solution
$u$ to Eq.~(\ref{wdmuDP}) with initial data $u_0$. Since
$H^3(\S)\subset C^2(\S)$ we find that
\bea\frac{\!\d}{\!\d t}\int_\S y^2\dx&=&2\int_\S
yy_t\dx\nonumber\\
&=&-2\int_\S uy_xy\dx-6\int_\S u_xy^2\dx-2\lambda\int_\S
y^2\dx\nonumber\\
&=&-5\int_\S u_xy^2\dx-2\lambda\int_\S y^2\dx.\label{WY2.3}\eea
If we assume $u_0\in H^4(\S)$ and use that $H^4(\S)\subset
C^3(\S)$, we can obtain
\bea\frac{\!\d}{\!\d t}\int_\S y_x^2\dx&=&2\int_\S
y_xy_{tx}\dx\nonumber\\
&=&-2\int_\S y_xy_{xx}u\dx-8\int_\S y_x^2u_x\dx-6\int_\S
yy_xu_{xx}\dx-2\lambda\int_\S y_x^2\dx\nonumber\\
&=&-7\int_\S y_x^2u_x\dx-2\lambda\int_\S
y_x^2\dx.\label{WY2.4}\eea
Adding (\ref{WY2.3}) and (\ref{WY2.4}) we get
\beq\frac{\!\d}{\!\d t}\norm{y}_{H^1}^2=-7\int_\S y_x^2u_x\dx-5\int_\S
u_xy^2\dx-2\lambda\norm{y}_{H^1}^2.\label{WY2.5}\eeq
Next we observe that (\ref{WY2.5}) also holds true for $u_0\in
H^3(\S)$: We approximate $u_0$ in $H^3(\S)$ by functions $u_0^n\in
H^4(\S)$, $n\geq 1$. Let $u^n=u^n(\cdot,u_0^n)$ be the solution of
(\ref{wdmuDP}) with initial data $u_0^n$. By Theorem~\ref{LWP} we
know that
$$u^n\in C\left([0,T_n);H^4(\S)\right)\cap C^1\left([0,T_n);H^3(\S)\right),\quad n\geq 1,$$
$$y^n=\mu(u^n)-u_{xx}^n\in C\left([0,T_n);H^2(\S)\right)\cap C^1\left([0,T_n);H^1(\S)\right),\quad n\geq 1,$$
$u^n\to u$ in $H^3(\S)$ and $T_n\to T$ as $n\to\infty$. Since
$u^n_0\in H^4(\S)$, we have
$$\frac{\!\d}{\!\d t}\int_\S(y_x^n)^2\dx=-7\int_\S (y_x^n)^2u_x^n\dx-2\lambda\int_\S (y_x^n)^2\dx.$$
Since $u_n\to u$ in $H^3(\S)$ it follows that $u_x^n\to u_x$ in
$L_{\infty}(\S)$ as $n\to\infty$. Note also that $y^n\to y$ in
$H^1(\S)$ and $y_x^n\to y_x$ in $L_2(\S)$ as $n\to\infty$. We
deduce that, as $n\to\infty$, \eqref{WY2.4} also holds for $u_0\in
H^3(\S)$. If $u_x$ is bounded from below on $[0,T)$, i.e.,
$u_x\geq -c$, where $c$ is a positive constant, then we can apply
Gronwall's inequality to \eqref{WY2.5} and have
$$\norm{y}_{H^1}^2\leq\norm{y_0}_{H^1}^2\exp((7c-2\lambda)t).$$
This shows that $\norm{u}_{H^3}$ does not blow up in finite time.
The converse direction follows from Sobolev's embedding theorem.
This completes the proof of our assertion.
\endproof
\rem
The above proof shows that if $u_x$ stays bounded, then $u$ also
persists in $H^3$. Thus Theorem \ref{ableitung} provides us with a
sufficient criterion for global existence, namely the boundedness
of $\norm{u_x(t,\cdot)}_{\infty}$ as $t$ approaches $T$ from
below.
\endrem
It is well known that the mean $\mu(u)$ of a solution $u(t,\cdot)$
of the $\muDP$ equation is conserved, i.e., $\mu(u_0)=\mu(u)$; see \cite{EKK10}. We
now show that the mean $\mu(u)$ of a solution of the weakly
dissipative $\muDP$ equation decreases exponentially as $t$
increases from zero. More precisely, we prove that the damping
constant is equal to the dissipation parameter $\lambda$.
\lem\label{conservation} Let $u_0\in H^3(\S)$ and denote by
$u(t,\cdot)$ the solution of \text{\rm(\ref{wdmuDP})} obtained in
Theorem \text{\rm\ref{LWP}}. Then the mean of $u$ satisfies
$$\mu(u)=\mu(u_0)e^{-\lambda t}$$
for $t\geq 0$ in the existence interval of $u$. In particular, if
$\mu(u_0)=0$, then the mean of the solution $u$ is conserved.
\endlem
\proof By differentiating under the integral sign and using
(\ref{eqn_rewritten}), we obtain that
\bea\frac{\!\d}{\!\d t}\mu(u)&=&\mu\left(-uu_x-3\mu(u)
\partial_x\Lambda_{\mu}^{-2}u-\lambda
u\right)\nonumber\\
&=&-\mu\left(\frac{1}{2}\partial_x\left(u^2\right)\right)-3\mu(u)\mu
\left(\partial_x\Lambda_{\mu}^{-2}u\right)-\lambda\mu(u),\eea
as long as the solution $u(t,\cdot)\in H^3(\S)$ exists. Hence
$$\frac{\!\d}{\!\d t}\mu(u)=-\lambda\mu(u)$$
from which the lemma follows.\endproof
With the help of Lemma \ref{conservation}, we are able to
establish the following blow-up setting. It is important to notice
that our result shows the blow-up of smooth initial data. A
corresponding result for $\muDP$ (the case $\lambda=0$) can be found in
\cite{LMT}.
\thm Assume that $0\neq u_0\in C^{\infty}(\S)$ has zero mean and
that there is $x^*\in\S$ satisfying
\bea\label{condition} 0<1+\frac{\lambda}{u_{0x}(x^*)}<1.\eea
Let $u$ be the corresponding solution of \eqref{wdmuDP}.
Then there is $0<\tau<\infty$ such that $\norm{u_x(t)}_{\infty}$
blows up as $t\to\tau$. In particular, the solution $u$ blows up
in the $H^3$-norm in finite time.
\endthm
\proof Differentiating equation (\ref{eqn_rewritten}) with respect
to $x$ and the identity $\partial_x^2\Lambda_{\mu}^{-2}=\mu-1$
(see Sect.~\ref{appendix}) yield
$$u_{tx}+uu_{xx}+u_x^2+\lambda u_x=3\mu(u)(u-\mu(u)).$$
By Lemma \ref{conservation}, it follows that the right hand side
equals zero. Again, we denote by $\phi$ the local flow of the
time-dependent vector field $u(t,\cdot)$, i.e.,
$\phi_t=u\circ\phi$. We set
$$w:=\frac{\phi_{tx}}{\phi_x}=u_x\circ\phi$$
and with
$$\phi_{ttx}=[(u_{tx}+uu_{xx}+u_x^2)\circ\phi]\,\phi_x$$
we obtain
$$w_{t}=\frac{\phi_{ttx}\phi_x-(\phi_{tx})^2}{\phi_x^2}=(u_{tx}+uu_{xx})\circ\phi$$
and hence
$$w_t+w^2+\lambda w=0.$$
With $\Gamma:=-\lambda<0$, we finally arrive at the logistic
equation
$$w_t=w(\Gamma-w)$$
and standard ODE techniques show that the solution is given by
$$w(t)=\frac{\Gamma}{1+\left(\frac{\Gamma}{w(0)}-1\right)e^{-\Gamma t}}.$$
Recall that $w(0)=u_{0x}(x)$. By our assumption on $u_0$, we can
find a point $x^*\in\S$ with
$$0<1+\frac{\lambda}{u_{0x}(x^*)}<1.$$
If we set
$$\tau=-\frac{1}{\lambda}\ln\left(1+\frac{\lambda}{u_{0x}(x^*)}\right),$$
it follows that the solution must blow up in the $H^{3}$-norm.
\endproof
\rem Condition (\ref{condition}) means that we can find $x^*\in\S$
such that
\begin{enumerate}
\item[$(i)$] $u_{0x}(x^*)<0$ and
\item[$(ii)$] $|u_{0x}(x^*)|>\lambda$.
\end{enumerate}
Since we assume $\mu(u_0)=0$, it follows that $u_0$ must change
sign. Since $u_0\in C^{\infty}(\S)$, $u_0$ has to change sign at
least twice and so it is always possible to find $x^*\in\S$
satisfying $(i)$. Our second condition says that the slope of
$u_0$ must decrease $\lambda$ in order to obtain blow-up: The
larger the dissipation given by $\lambda$, the larger must
$|u_{0x}|$ be locally in order to obtain a blow-up. So
$(\ref{condition})$ is a non-trivial common condition for
$\lambda$ and $u_0$ in our blow-up setting.
\endrem
The following lemma is similar to Lemma 2.2.\ in \cite{EWY}.
Furthermore, we see that as $\lambda\to 0$, we obtain the
conservation of the quantity $(y\circ\phi)\phi_x^3$, which is
explained in \cite{EK,LMT} for the DP and the $\mu$DP.
\lem\label{conservation2} Let $u_0\in H^3(\S)$ and let $T>0$ be
the maximal existence time of the corresponding solution $u(t,x)$
according to Theorem \text{\rm\ref{LWP}}. Let $\phi$ be the local
flow of $u$ according to Lemma~\ref{lem_flow}. Then we have
$$y(t,\phi(t,x))\phi_x^3(t,x)=y_0(x)e^{-\lambda t}.$$
\proof An easy calculation shows that the function
$$[0,T)\mapsto\R,\quad t\mapsto e^{\lambda t}y(t,\phi(t,x))\phi_x^3(t,x)$$
is constant. Using $\phi(0)=\id$ and $\phi_x(0)=1$, we are done.
\endproof
Finally, we come to the following global well-posedness result.
Note that our assumptions on the initial condition $u_0$ are quite
similar to the ones in Theorem~5.4.\ in \cite{LMT}.
\thm\label{GWP} Assume that $u_0\in H^3(\S)$ has positive mean and
satisfies the condition $\Lambda_{\mu}^2u_0\geq 0$. Then the
Cauchy problem \text{\rm(\ref{wdmuDP})} has a unique global
solution in $C([0,\infty),H^3(\S))\cap C^1([0,\infty),H^{2}(\S))$.
\proof Let $u(t,\cdot)\in H^{3}(\S)$, $t\in[0,T)$, denote the
solution of (\ref{wdmuDP}) obtained in Theorem~\ref{LWP}.
According to Theorem \ref{ableitung}, we only have to show that
$\norm{u_x(t,\cdot)}_{\infty}$ stays bounded as $t$ approaches $T$
from below. Note that, for any periodic function $w$,
differentiating formula (\ref{inverse}) with $f=\Lambda_\mu^2w$ yields
$$\norm{\partial_xw}_{\infty}\leq C\norm{\Lambda_{\mu}^2w}_{L_1},$$
with a constant $C\geq 0$. Now Lemma \ref{conservation2}
and the assumption $\Lambda_{\mu}^2u_0\geq 0$ imply that
$$\norm{\Lambda_{\mu}^2u}_{L_1}=\mu\left(\Lambda_{\mu}^2u\right).$$
Using Lemma \ref{conservation}, we have the estimate
$$\norm{\partial_xu(t,\cdot)}_{\infty}\leq C\int_0^1\Lambda_{\mu}^2u\dx=C\mu(u)\leq C\mu(u_0)<\infty,$$
from which the indefinite persistence of the solution $u$ follows.
\endproof
It is clear that Theorem \ref{GWP} also holds if
$\Lambda_{\mu}^2u_0\leq 0$ and $\mu(u_0)<0$.
\section{Appendix}\label{appendix}
We denote by $H^k=H^k(\S)$, $k\geq 0$, the Sobolev space of
periodic functions. If $k\in\N_0$, $H^k$ is the space of all
$L_2(\S)$-functions $f$ with square integrable distributional
derivatives up to the order $k$, $\partial_x^jf\in L_2(\S)$,
$j=0,\ldots,k$. Endowed with the norm
$$\norm{f}_k^2=\sum_{j=0}^k\int_{\S}(\partial_x^jf)^2(x)\dx=\sum_{i=0}^k\ska{\partial_x^jf}
{\partial_x^jf}_{L_2(\S)}=\sum_{j=0}^k\norm{\partial_x^jf}_{L_2(\S)}^2,$$
the spaces $H^k$ become Hilbert spaces. Note that we have
$H^0=L_2(\S)$. To define the spaces $H^k$ for general $k\geq 0$,
we make use of the fact that the Fourier transform $\mathscr F$ maps any
square integrable function $f$ on $\S$ to its Fourier series
$(\hat f(n))_{n\in\Z}$ so that $f(x)=\sum_{n\in\Z}\hat f(n)e^{2\pi\text{i}nx}$. The space $H^k$ consists of all
$f\in L_2(\S)$ with the property that the quadratic form $\ska{Q^{2k}f}{f}_{L_2(\S)}$ has a finite value,
where $Q=(1-\partial_x^2)^{1/2}$ is the elliptic pseudo-differential operator
with the symbol $(1+4\pi^2n^2)^{1/2}$, i.e.,
$$(\mathscr F(Q^k f))(n)=(1+4\pi^2n^2)^{k/2}\hat f(n).$$
We thus have
$$H^k(\S):=\set{f\in L_2(\S)}{\norm{f}_k^2=\sum_{n\in\Z}\left|(\mathscr F(Q^kf))(n)\right|^2<\infty}.$$
It is
easy to check that the operator $\Lambda_{\mu}^2=\mu-\partial_x^2$
has the inverse
\bea\label{inverse}(\Lambda_{\mu}^{-2}f)(x)&=&\left(\frac{1}{2}x^2-\frac{1}{2}x+
\frac{13}{12}\right)\int_0^1f(a)\d a+\left(x-\frac{1}{2}\right)
\int_0^1\int_0^af(b)\d b\d a\nonumber\\
&&-\int_0^x\int_0^af(b)\d b\d a+\int_0^1\int_0^a\int_0^bf(c)\d c\d
b\d a.\eea
To obtain Green's function $g(x-x')$ for $\Lambda_\mu^{-2}$,
we observe that applying $\Lambda_\mu^2$ to
$$g(x)=\frac{1}{2}x^2-\frac{1}{2}|x|+\frac{13}{12}$$
gives the delta distribution. Hence
$$(\Lambda_\mu^{-2}f)(x)=\int_0^1g(x-x')f(x')\d x'.$$
Moreover, we see that $[\partial_x,\Lambda_{\mu}^{-2}]=0$ and
$\partial_x^2\Lambda^{-2}_{\mu}=\mu-1$. It is also easy to verify
that $\Lambda_\mu^2\colon H^k\to H^{k-2}$, $k\geq 2$, is a topological
isomorphism: For any $f\in H^k$ we have
\beq\norm{\Lambda_\mu^2f}_{k-2}^2=\norm{\hat f(0)+\sum_{n\neq
0}\hat f(n)4\pi^2 n^2e^{2\pi\text{i}nx}}_{k-2}^2\leq
2\sum_{n\in\Z}(1+4\pi^2n^2)^{k}\left|\hat
f(n)\right|^2=2\norm{f}_k^2;\nonumber\eeq
together with (\ref{inverse}) the open mapping theorem achieves
the desired result.
\end{document}